\documentclass[aps,prb,twocolumn,showpacs,preprintnumbers,superscriptaddress,amsmath,amssymb,longbibliography,floatfix]{revtex4-2}
\usepackage{graphicx}% Include figure files
\usepackage{dcolumn}% Align table columns on decimal point
\usepackage{bm}% bold math
\usepackage{amsfonts}
\usepackage{amsmath}
\usepackage{amssymb}
\usepackage{color}
\usepackage{enumerate}
\usepackage{colortbl}
\usepackage[utf8]{inputenc}
\newcolumntype{P}[1]{>{\centering\arraybackslash}p{#1}}

\DeclareUnicodeCharacter{2212}{-}
\DeclareUnicodeCharacter{0301}{´}

\usepackage[usenames,dvipsnames]{xcolor}

\begin{document}

\title{Interfacial and thickness effects in La$_{2/3}$Sr$_{1/3}$MnO$_3$/YBa$_2$Cu$_3$O$_{7}$ superlattices}% Force line breaks with \\

\author{V. A. M. Lima}
\affiliation{Departamento de Física, Universidade Federal de Minas Gerais, C.P. 702, 30123-970 Belo Horizonte, Minas Gerais, Brazil}
\author{M. C. O. Aguiar}
\affiliation{Departamento de Física, Universidade Federal de Minas Gerais, C.P. 702, 30123-970 Belo Horizonte, Minas Gerais, Brazil}
\author{N. C. Plumb}
\affiliation{Swiss Light Source, Paul Scherrer Institut,  CH-5232 Villigen, Switzerland}
\author{M. Radovic}
\affiliation{Swiss Light Source, Paul Scherrer Institut,  CH-5232 Villigen, Switzerland}
\author{W. H. Brito}
\affiliation{Departamento de Física, Universidade Federal de Minas Gerais, C.P. 702, 30123-970 Belo Horizonte, Minas Gerais, Brazil}

\date{\today}% It is always \today, today,

\begin{abstract}

Superlattices of correlated oxides have been used to explore interfacial effects and to achieve additional control over the physical properties of individual constituents. In this work, we present a first-principles perspective of the strain and thickness effects in La$_{2/3}$Sr$_{1/3}$MnO$_3$/YBa$_2$Cu$_3$O$_{7}$ (LSMO/YBCO) superlattices. Our findings indicate that the presence of epitaxial strain and LSMO leads to a reduction of buckling parameters of the interfacial CuO$_2$ planes, as well as the transfer of electrons from LSMO to YBCO. In addition, the change in Cu-3$d$ valence is slightly dependent on the LSMO layer thickness.
More interestingly, the in-plane ferromagnetic ground state within the CuO$_2$ planes near the interface is induced due to the local moments centered at the copper atoms. These local moments are decoupled from the charge transfer and, according to our calculations, appear mainly due to the Mn $3d$-O $2p$-Cu $3d$ hybridization being restricted to the interfacial region. 
The induced net magnetic ordering in interfacial copper atoms may have implications in the control of the superconducting state in the LSMO/YBCO superlattices.
 
\end{abstract}

\maketitle

%\tableofcontents

\section{\label{sec:int}Introduction} 

Atomic precision growth methods and advanced characterization techniques have paved the way for the study of transition metal oxides (TMO's) heterostructures and superlattices. In 2004, most of the scientific interest in TMO's heterostructures was triggered by the discovery of a two-dimensional electron gas (2DEG) at SrTiO$_3$/LaAlO$_3$ interface~\cite{sup_gas_2d}. Since then, numerous studies have been carried out to elucidate the physical mechanisms underlying the properties of the 2DEG, its superconducting phase~\cite{SC_STOLAO}, and interfacial ferromagnetism~\cite{ferromagnetism_STOLAO}.
Moreover, interfacial and proximity effects have also been used to explore the emergent properties of cuprate high-temperature superconductors, where a complex interplay among competing orders takes place.

The intriguing balance between the competing orders in the cuprates can be affected by the hole density, the number of CuO$_2$ layers in the structure, epitaxial strain, interfacial coupling, and proximity effects. According to the early work of Chakhalian and coworkers~\cite{dead_layer}, the interfacial coupling between La$_{2/3}$Ca$_{1/3}$MnO$_3$ (LCMO) and YBa$_2$Cu$_3$O$_7$ (YBCO) gives rise to small uncompensated magnetic moments at Cu atoms near the interface, mainly due to the interactions with underneath Mn atoms. Later on, the same group also found an orbital reconstruction in LCMO/YBCO heterostructures, where the Cu-3$d_{z^{2}}$ states near the interface become partially occupied~\cite{xas_chg_transf} due to the hybridization with Mn-3$d_{z^2}$ orbitals. The unusual magnetic profile of the LCMO/YBCO interface was also addressed by neutron experiments~\cite{Stahn_PRB} and density functional theory (DFT) calculations~\cite{dead_layer_teo,yang2009electronic}, which pointed out the formation of a magnetic "dead layer", associated with the antiferromagnetic coupling between MnO$_2$ layers close to the interface. More recently, x-ray scattering measurements indicated that the coupling between Cu and Mn atoms is sensitive to the interfacial terminations, where antiferromagnetic and ferromagnetic couplings can be realized~\cite{selective_interlayer_coup}. 

To understand the complex physics that takes place at interfaces between superconductors and ferromagnetic oxides, it is also important to clarify how the charge transfer evolves across the interface. Overall, previous theoretical~\cite{teo_elec_dop,Gonzalez_2008} and experimental~\cite{xas_chg_transf} works have indicated that electrons are transferred from the manganite to the cuprate in LCMO/YBCO and La$_{2/3}$Sr$_{1/3}$MnO$_3$ (LSMO)/YBCO interfaces, leading to a reduction of Mn-$e_{g}$ occupancy. Further experimental studies have indicated that in LCMO/YBCO heterostructures, the charge transfer occurs only within a few atomic planes from the interface~\cite{Chien2013}. More interestingly, XAS experiments~\cite{Tra_APL2017} have shown that the superconducting and magnetic properties of LCMO/YBCO heterostructures are sensitive to the manganite termination, which also affects the charge transfer between the two oxides. This intriguing proximity effect and charge transfer can also lead to superconductor to Mott insulator transitions in LCMO/YBCO superlattices, where the transition is controlled by the LCMO thickness~\cite{sic_ybco_lsmo}. More recently, these heterostructures have been used to study the interplay of charge-density wave (CDW) and superconductivity~\cite{Frano2016}, to disclose a new type of CDW ordering associated with Cu-3$d_{z^2}$ orbitals~\cite{Gaina2021}, to induce spin-triplet superconductivity~\cite{Kumawat2023}, and to investigate the Josephson coupling in LSMO/YBCO/LSMO trilayer junctions~\cite{Sanchez-Manzano2022}.  

Motivated by the experimental and theoretical findings mentioned above, we have performed an extensive theoretical investigation of LSMO/YBCO superlattices by means of first-principles calculations. Since LSMO and LCMO are ferromagnetic half-metallic manganese oxides~\cite{mixed_valence_manganites} and the studies report a similar qualitative behavior for both LSMO/YBCO and LCMO/YBCO superlattices and heterostructures,
we expect that our results can be extended to LCMO interfaces. We emphasize that the details of structural distortions, charge transfer, interfacial magnetism, and confinement presented in LSMO/YBCO superlattices remain unclear and are of great importance for a better understanding of the interface physics associated with strongly correlated oxides. Our goal is to better understand how the change in these parameters affects the cuprate and the manganite properties. In addition, our work is complementary to previous first principles studies of manganite/cuprate heterostructures, which were able to reproduce some of the physical features observed experimentally~\cite{dead_layer_teo,teo_elec_dop}.

Our findings indicate that epitaxial strain and the presence of LSMO suppresses the buckling distances and angles of the interfacial CuO$_2$ layers. These structural distortions are accompanied by a transfer of electrons from LSMO to YBCO, which can be explained in terms of electronegativity difference. We also find induced local moments centered at the interfacial Cu atoms, which, in turn, are decoupled from the charge transfer. These local moments lead to an in-plane ferromagnetic ordering and appear mainly due to the Mn 3d-O 2p-Cu 3d hybridization.

The paper is organized in four sections.  
In Sec.~\ref{sec:mod}, we present our proposed structural models as well as the employed theoretical approximations. The relaxed structures and structural parameters are discussed in Sec.~\ref{subsec:str}, while the calculated interfacial charge transfer and magnetism are discussed in Secs.~\ref{subsec:chg} and~\ref{subsec:mag}, respectively. 
The hybridization between the interfacial Mn 3d-O 2p-Cu 3d states is discussed in Sec.~\ref{int_hyb}. 
Our conclusions are summarized in Sec.~\ref{sec:conc}.

\section{\label{sec:mod}Models and Methods}

\begin{figure*}
\includegraphics[scale=0.55]{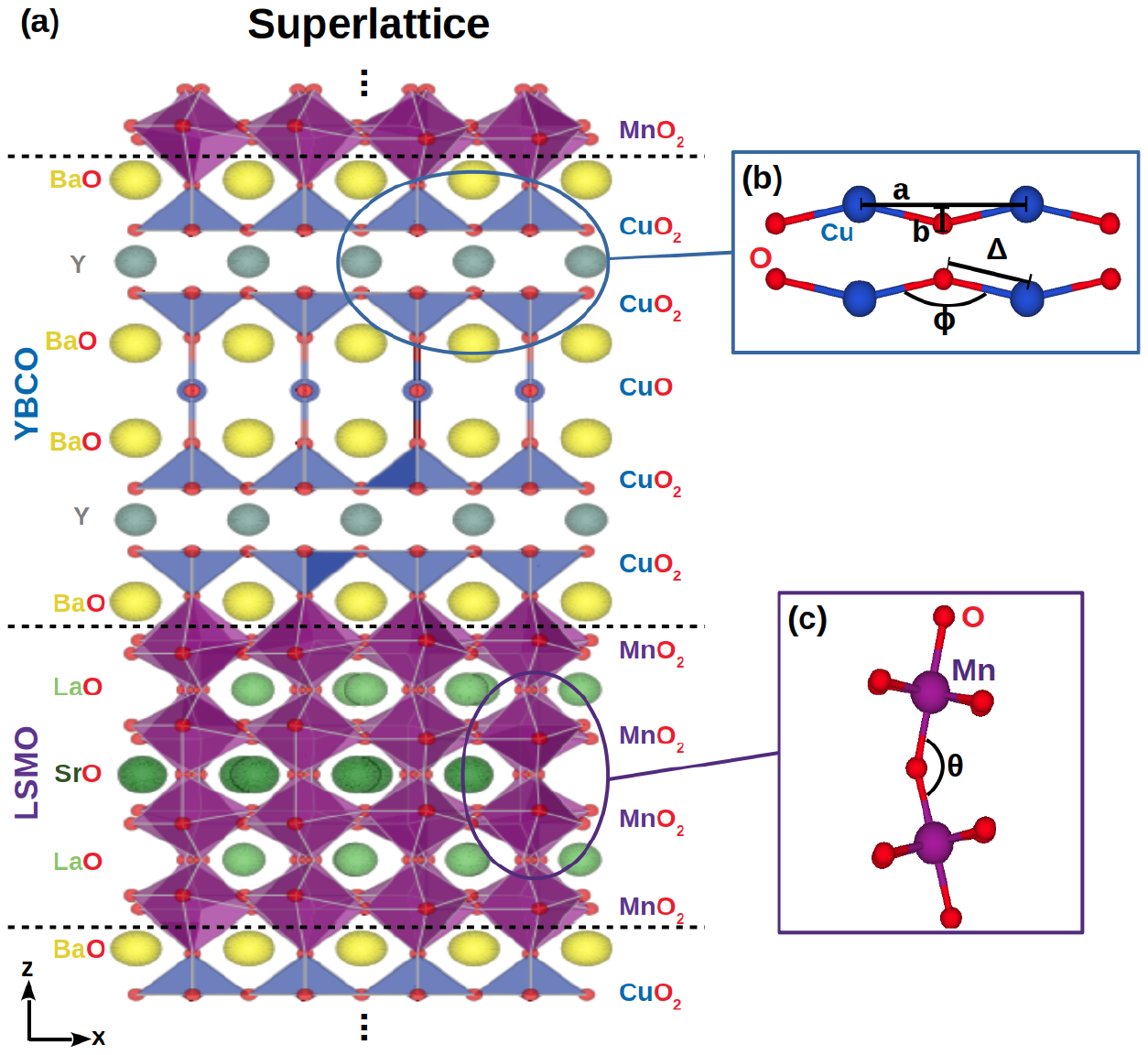}% Here is how to import EPS art
\caption{(a) Structural model of a superlattice with plane-terminated interfaces. (b) The structure of the buckling distance $b$, the TM-O-TM (TM =Cu, Mn) bond angle $\phi$, the TM-O (planar) distance $\Delta$, and the TM-TM in-plane distance $a$. (c) The structure of the out-of-plane TM-O-TM (TM = Cu, Mn) bond angle $\theta$.}
\label{fig:1}
\end{figure*}

To study the physical properties of (001)-oriented LSMO/YBCO superlattices we constructed La$_{2/3}$Sr$_{1/3}$MnO$_3$ ($m$ u.c.)/YBa$_2$Cu$_3$O$_7$ ($n$ u.c.) structural models, with $m$ = 3, 6 and $n$ = 2, 4, using tetragonal supercells with MnO$_{2}$-BaO interface termination (plane terminated) (see Fig.~\ref{fig:1}(a)).
The LSMO thicknesses were chosen to allow the use of more homogeneous Sr doping. Regarding the YBCO, we model the two YBCO layers with different numbers of unit cells to understand the effects of changing the thickness.
We also used $\sqrt{2}a\times\sqrt{2}a$ in-plane rotated supercells to take into account the MnO$_6$ octahedra distortions, which are essential for describing the LSMO properties~\cite{mixed_valence_manganites}. To reduce complexity, Sr-doping was considered by replacing La atoms with Sr atoms in the inner layers (away from the interface) of LaMnO$_3$ (LMO).

The structural parameters addressed in our work are illustrated in Figs.~\ref{fig:1} (b) and (c). The buckling distance $b$ measures how distant the oxygen atoms are from a perfectly flat TMO$_2$ plane (TM for transition metal), $\phi$ corresponds to the in-plane TM-O-TM (TM =Cu, Mn) bond angle, and $\Delta$ represents the TM-O (planar) distance.
From now on, the LSMO/YBCO superlattices will be labeled as (LSMO)$_m$/(YBCO)$_n$ for simplicity. We also constructed LaMnO$_3$ (2 u.c.)/YBa$_2$Cu$_3$O$_7$ (2 u.c.) and LaMnO$_3$ (3 u.c.)/YBa$_2$Cu$_3$O$_7$ (2 u.c.) superlattices, labeled as (LMO)$_2$/(YBCO)$_2$ and (LMO)$_3$/(YBCO)$_2$, respectively, to compare with (LSMO)$_m$/(YBCO)$_n$ structures and analyze the Sr-doping effects. To investigate the effects of periodicity in $z$-axis direction, we also built a LaMnO$_3$ (2 u.c.)/YBa$_2$Cu$_3$O$_7$ (2 u.c.) heterostructure model (see Appendix~\ref{het}). In all the structures considered, there are two types of Cu atoms: the ones belonging to the CuO$_2$ planes, labeled as Cu(p$i$), and the ones belonging to the CuO chains, labeled as Cu(c$j$). The $i$ and $j$ indexes mark the $z$-axis position counting from the interface.
For all the configurations, we imposed an epitaxial strain induced by SrTiO$_3$ (STO)(001), a common substrate used when experimentally growing oxide heterostructures, by fixing the in-plane lattice parameters to $a_{STO}$ = 3.905 \AA{}.

We performed spin-polarized DFT+U~\cite{dftu_lich} calculations within the Perdew-Burke-Ernzerhof (PBE) generalized gradient approximation~\cite{pbe} as implemented in the Vienna $\textit{Ab initio}$ Simulation Package (VASP)~\cite{vasp1,vasp2,vasp3}.
We used projector augmented waves (PAW) pseudopotentials~\cite{paw} and a plane-wave energy cutoff of 450 eV.
PAW-PBE pseudopotentials in VASP have great reliability for a significant group of materials and were extensively tested~\cite{pseudo_test}.
Additional electronic interactions, at the static mean-field level, were considered using a local electronic interaction $\textit{U}$ = 5 eV and a Hund's coupling $\textit{J}$ = 1 eV for the Mn-3$d$ orbitals.
The atomic positions, as well as the superlattices sizes in $z$-axis direction, were relaxed using a 4 $\times$ 4 $\times$ 1 \textit{k}-point mesh until the forces on each atom were less than 0.01 eV/\AA{}. In all structures, the density of states (DOS) was calculated using a 12 $\times$ 12 $\times$ 3 \textit{k}-point mesh. 

\section{\label{sec:res}Results and discussions}

\subsection{\label{subsec:str}Structural distortions and buckling suppression}

\begin{figure*}
\includegraphics[scale=0.4]{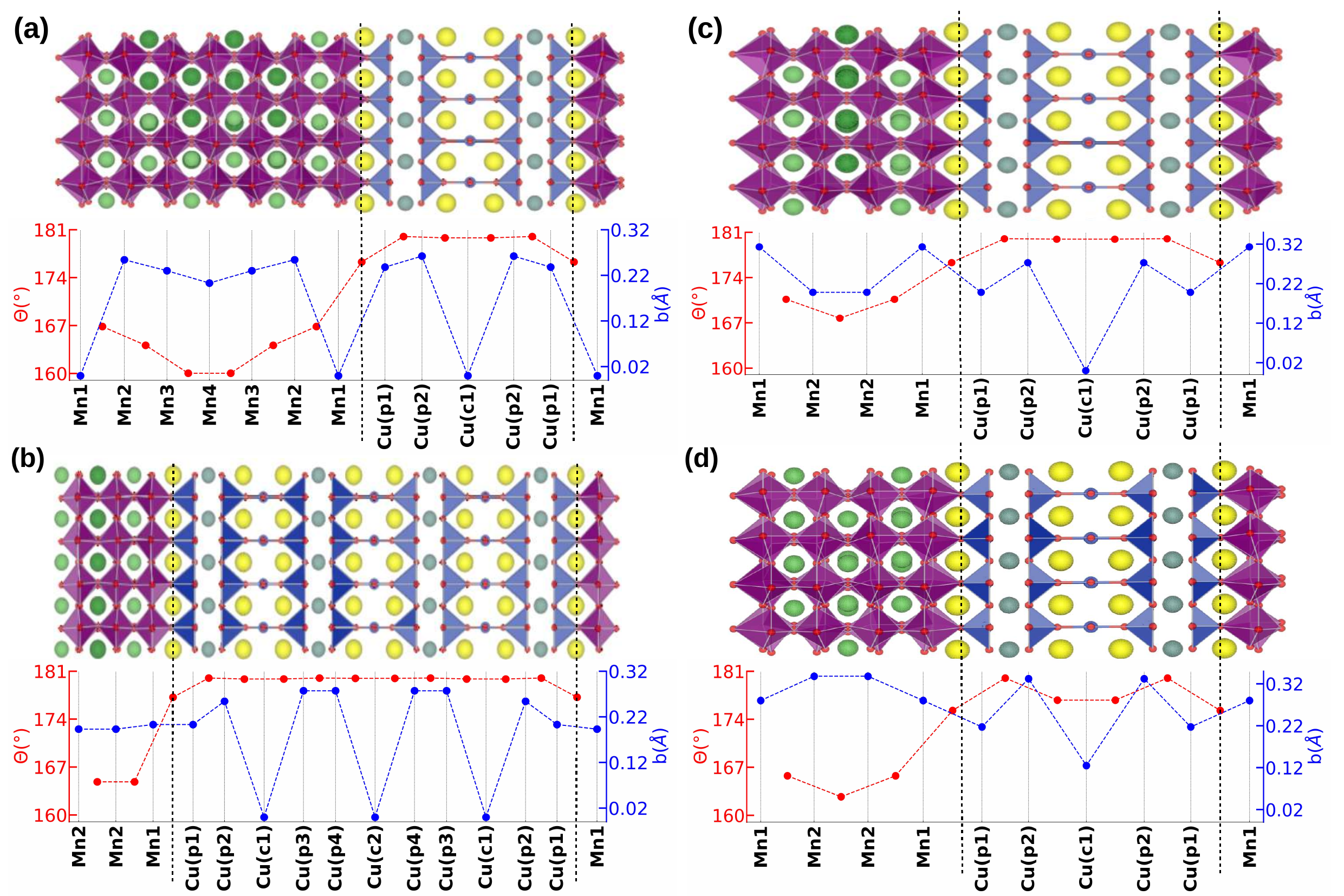}% Here is how to import EPS art
\caption{Relaxed structures, out-of-plane TM-O-TM bond angles $\theta$ (in red) and in-plane buckling $b$ (in blue) for (a) (LSMO)$_6$/(YBCO)$_2$, (b) (LSMO)$_3$/(YBCO)$_4$, (c) (LSMO)$_3$/(YBCO)$_2$ and (d) (LMO)$_3$/(YBCO)$_2$ superlattices. In Mn$_i$ $i$ indicates the $z$-axis position counting from the interface, whose location is given by the black dashed lines.}
\label{fig:2}
\end{figure*}

In Figs.~\ref{fig:2}(a) and (b), we display the relaxed structures of (LSMO)$_6$/(YBCO)$_2$ and (LSMO)$_3$/(YBCO)$_4$ superlattices, respectively. In addition, we show the calculated out-of-plane TM-O-TM (TM =Cu, Mn) bond angles ($\theta$) (in red) and buckling distances ($b$) (in blue).
We observe that atoms near the interface present different distortion patterns compared to our relaxed bulk (isolated) YBCO and LSMO. 
In particular, we find that $b$ is slightly reduced for the CuO$_2$ planes at the interface for both superlattices. 
In (LSMO)$_6$/(YBCO)$_2$, we find a reduction of $b$ of around 0.11\AA{}, in comparison with bulk YBCO ($b_{bulk}=0.35$\AA{}), for the interfacial CuO$_2$ planes. The same feature is observed in (LSMO)$_3$/(YBCO)$_4$ (Fig. \ref{fig:2}(b)), where the $b$ is reduced by 0.15\AA{} at the interface. On the LSMO side, one can notice a stronger suppression of the Mn-O buckling distance for (LSMO)$_6$/(YBCO)$_2$, in comparison with bulk LSMO, and an increase of the Mn2-O-Mn1 bond angle in both structures (in red).

We mention that, according to early studies of Kambee and Ishii~\cite{KAMBE2000555}, the buckling distances are connected with the superconducting critical temperature ($T_c$): higher transition temperatures were found for cuprates with smaller buckling distances $b$. This can be understood within a magnetic mechanism for Cooper pairing where the superexchange coupling $J_x \propto cos^2 \phi$, where $\phi$ is the buckling angle~\cite{Keren_2009} (see Fig.~\ref{fig:1}(b)). According to our results, the reduction of $b$ is accompanied by an average reduction of $\phi$ of around 1.56$^\circ$ in the superlattices with respect to bulk values. Therefore, according to our findings, the epitaxial strain and interface structural distortions would enhance the superconducting phase of CuO$_2$ planes near the interface. 

As pointed out in Ref.~\onlinecite{mixed_valence_manganites}, the magnetic ordering in LSMO is directly connected to the change in the Mn-O-Mn bond angle $\theta$ upon doping, going from an AFM out-of-plane coupling ($\theta\sim155^\circ$ in LaMnO$_3$) to a FM coupling ($\theta\sim160^\circ$ in La$_{2/3}$Sr$_{1/3}$MnO$_3$), followed by an AFM coupling ($\theta\sim170^\circ$ in La$_{1/10}$Sr$_{9/10}$MnO$_3$). As can be seen in Fig.~\ref{fig:2} (red lines), near the interface $\theta$ approaches values corresponding to large doping, suggesting an AFM coupling of MnO$_2$ layers (corresponding to a magnetic "dead layer"). On the YBCO side, $\theta$ is only affected at the interface, where the Mn-O-Cu angle is 176.28$^\circ$ for (LSMO)$_6$/(YBCO)$_2$ and 177.19$^\circ$ for (LSMO)$_3$/(YBCO)$_4$. For (LSMO)$_3$/(YBCO)$_2$ (see now Fig.~\ref{fig:2}(c)), we find $\theta = $  177.03$^\circ$, indicating that the $m/n$ ratio is one of the control parameters on the interfacial distortions. It is important to note that the angle $\theta$ would be smaller in YBCO ultrathin films on thicker LSMO substrates. We mention that the Cu-O-Mn out-of-plane interface bond length remains almost the same in (LSMO)$_6$/(YBCO)$_2$, (LSMO)$_3$/(YBCO)$_4$ and (LSMO)$_3$/(YBCO)$_2$ ($\sim$ 4.20 \AA{}).

The results for the heterostructure (see Appendix~\ref{het}) suggests that the buckling distance $b$ is slightly reduced for the CuO$_2$ planes at the interface, in comparison with the bulk values, accompanied by an average reduction of the buckling angle $\phi$ and that the behavior of $\theta$ near the interface also suggests the formation of a magnetic ``dead layer".
 
Finally, we investigate how the Sr-doping affects the key structural distortions by comparing the relaxed structures of (LSMO)$_3$/(YBCO)$_2$ and (LMO)$_3$/(YBCO)$_2$ (Figs.~\ref{fig:2}(c) and (d)). We observe that the YBCO structure is more affected when using LMO instead of LSMO, which can be connected to bulk LMO exhibiting larger Mn-O-Mn bond angle distortions than LSMO. This results in $\theta$ along the whole structure becoming smaller in (LMO)$_3$/(YBCO)$_2$ than in (LSMO)$_3$/(YBCO)$_2$. We also find that the Cu-O bond length of the chain is larger in (LMO)$_3$/(YBCO)$_2$ (1.977 \AA{}) than in (LSMO)$_3$/(YBCO)$_2$ (1.953 \AA{}), with the latter being the same as in bulk YBCO. Together with the increase of these out-of-plane distortions, the bond length $\Delta$ also increases in the CuO$_2$ planes when using LMO (not shown), indicating a larger tilting on the CuO$_2$ tetrahedral than in (LSMO)$_3$/(YBCO)$_2$. However, the Cu-O-Mn out-of-plane interface bond length for (LMO)$_3$/(YBCO)$_2$ is $\sim2\%$ larger than for (LSMO)$_3$/(YBCO)$_2$.

We emphasize that when the CuO$_2$ plane is flat (small $b$), $T_c$ is enhanced, the large buckling values tend to localize the conducting holes in underdoped cuprates, while decreasing the superconducting strength~\cite{KAMBE2000555}. By looking only at the structural distortions, our findings suggest an enhancement of $T_c$ in all structures, which is indeed in disagreement with the experimental findings~\cite{sic_ybco_lsmo}. This shows that the structural distortions can not explain the suppression of superconductivity in the superlattices and heterostructures~\cite{APL_Sefrioui,Dybko_2013}. On the other hand, previous works have reported a reduction in $T_c$ for bulk YBCO~\cite{strain_tc} induced by $z$-axis compressive strain effects. We display the key structural parameters in Appendix~\ref{st_tc}.

\subsection{\label{subsec:chg}Charge transfer}

\begin{figure*}
\includegraphics[scale=0.5]{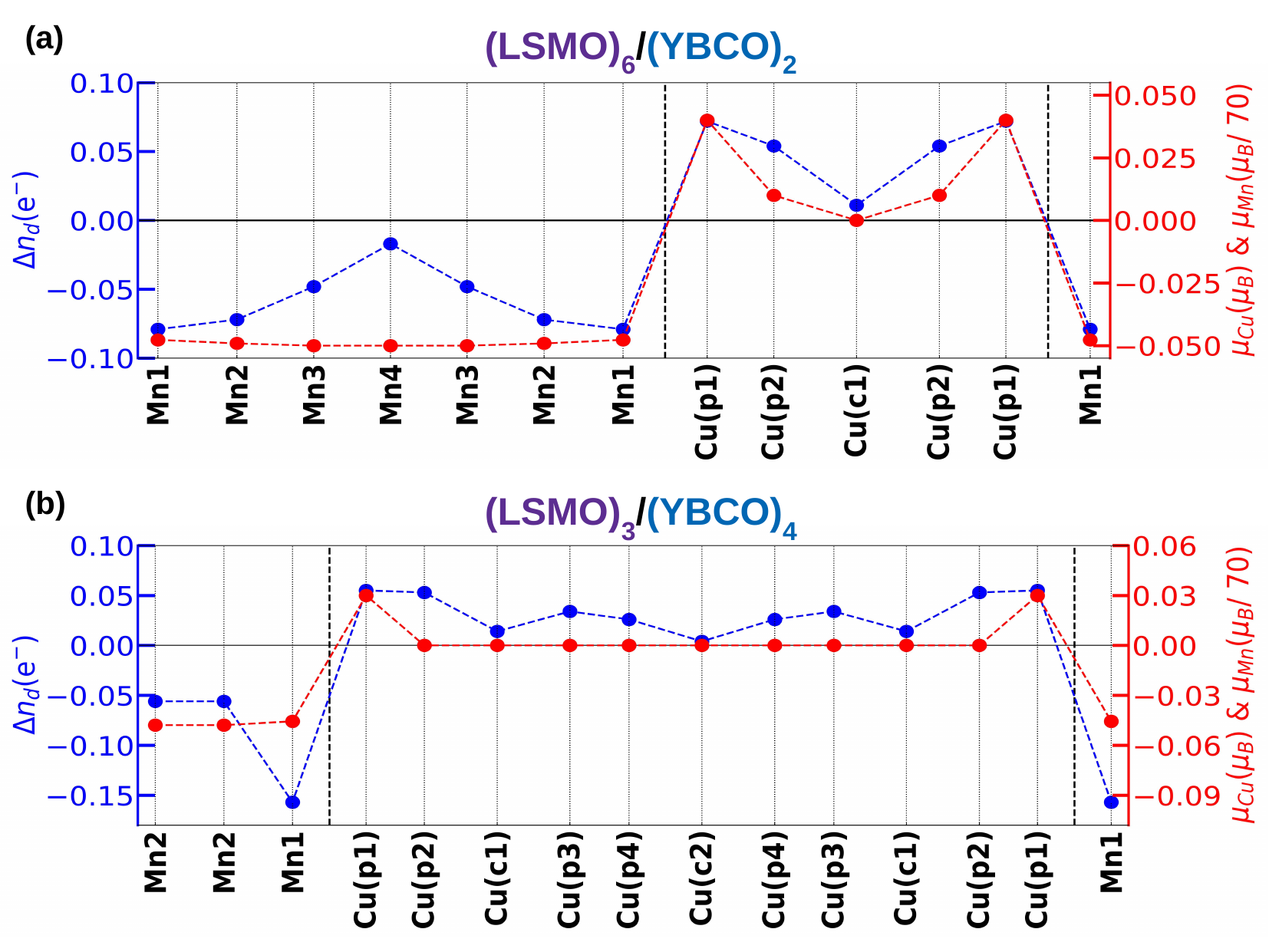}% Here is how to import EPS art
\caption{$3d$-orbital charge difference $\Delta n_d$ (blue) and magnetic moments $\mu$ (red) for the TM along the structure for (a) (LSMO)$_6$/(YBCO)$_2$ and (b) (LSMO)$_3$/(YBCO)$_4$ superlattices. The dashed black lines indicate the interfaces. For better visualization, the Mn moments were reduced by a factor of 70. $\Delta n_d$ was calculated using the difference between the interface and bulk occupations.}
\label{fig:4}
\end{figure*}

Another important feature of oxide interfaces is the charge transfer between the constituents, which can give rise to different $d$ shell occupancies from those observed in the bulk counterparts. 
The charge transfer driving mechanisms have been addressed in several previous works~\cite{Chen_2017, Zhong_PRX_2017, Beck_PRM2019}, where the band alignment of O-$2p$ and TM-$3d$ states between the oxides forming the superlattice and the electronegativity difference play a central role. According to Chen and Millis~\cite{Chen_2017}, the energy separation between the TM-3$d$ and O-2$p$ states is larger in bulk LSMO than in bulk YBCO. Therefore, the electronegativity difference of our system indicates that LSMO would transfer electrons to YBCO. To investigate this hypothesis, we study the charge transfer within our models and how the 3$d$ occupancies change across the interfaces. In particular, we evaluate integrals of the projected density of states up to the Fermi energy for the distinct Cu and Mn atoms presented in our structures and compare them with the occupation values of bulk YBCO and LSMO.

In Fig. \ref{fig:4} we show the variation of 3$d$ occupancies (in blue), $\Delta n_d = n_{SL} - n_{bulk}$, for Mn and Cu atoms for the relaxed superlattices (SL's) in comparison with the bulk YBCO occupation values. For (LSMO)$_6$/(YBCO)$_2$ (Fig. \ref{fig:4}(a)), we find an increase of Cu-3$d$ occupancies for the CuO$_2$ planes and the CuO chain, accompanied by a reduction of Mn-3$d$ occupancies. In particular, we find that Cu-3$d$ occupancy of interfacial CuO$_2$ planes increases by 0.072 electrons, in agreement with previous experimental studies~\cite{Chien2013}. The 3$d$ occupancy of the inner CuO chain, in turn, increases by around 0.011 electrons.
Similar charge transfer profiles are observed for (LSMO)$_3$/(YBCO)$_4$ (Fig. \ref{fig:4}(b)), where the Cu-3$d$ interfacial occupancy increases by 0.055 electrons. This increase of Cu-3$d$ occupancies should drive the CuO$_2$ planes away from the optimal regime towards the antiferromagnetic phase. For the heterostructure (see Appendix~\ref{het}), the Cu-3$d$ interfacial occupancies also increase.

\begin{figure*}
\includegraphics[scale=0.5]{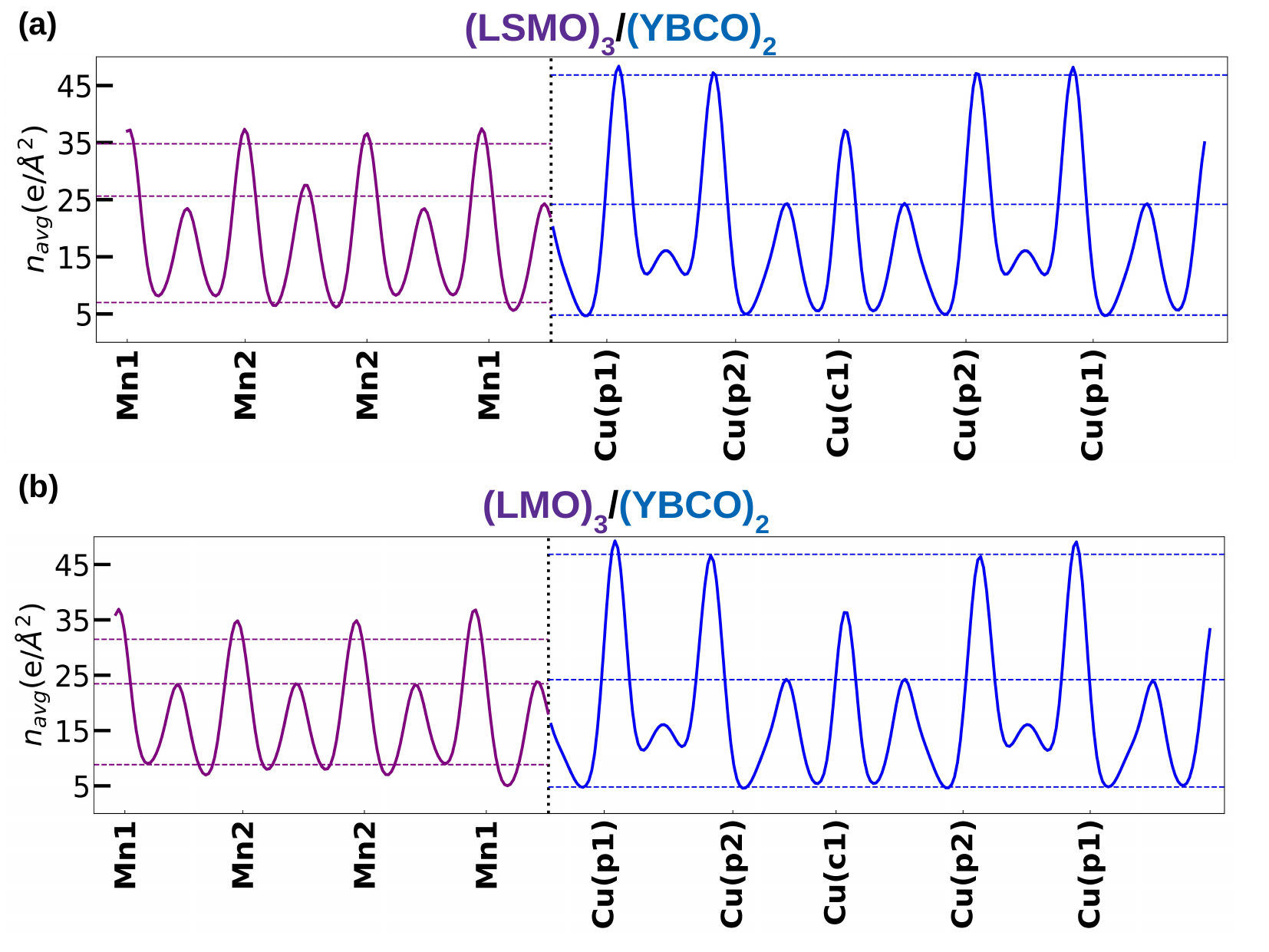}% Here is how to import EPS art
\caption{Average planar charge density $n_{avg}$ for (a) (LSMO)$_3$/(YBCO)$_2$ and (b) (LMO)$_3$/(YBCO)$_2$ superlattices. The dashed black lines indicate the interface; the purple lines indicate the bulk LSMO (in (a)) and LMO (in (b)) average planar charge density corresponding to the first and the second maxima as well as the minimum value, while the blue lines indicate the same for YBCO.}
\label{fig:5}
\end{figure*}

The effects of Sr-doping on the charge transfer can be analyzed by comparing (LSMO)$_3$/(YBCO)$_2$ and (LMO)$_3$/(YBCO)$_2$ average planar charge densities, as shown in Fig. \ref{fig:5}.
In the inner MnO$_2$ layers of (LMO)$_3$/(YBCO)$_2$, the charge average is smaller than in (LSMO)$_3$/(YBCO)$_2$, but at the interface, the charge profiles are similar (purple lines in Figs. \ref{fig:5}(a) and (b)). We find the highest peak intensity for the LMO interface. In LSMO, the Sr has a 2+ valence rather than a 3+ valence, which is associated with La~\cite{mixed_valence_manganites}. This reduces the Mn occupation, changing it from 3d$^4$ to 3d$^3$ in a proportion of $\frac{\text{\# La atoms}}{\text{\# Sr atoms}}=x$.
In addition, one can see in Fig. \ref{fig:5} that the charge of the LaO and BaO semiconducting layers is also affected by the interface, with a reduction of the maximum charge for the ones close to the interface; it can explain the electron doping in the first MnO$_2$ plane present in both figures. As a result, the charge imbalance at the interface, caused by the MnO$_2$-BaO-CuO$_2$ termination, makes a "bridge" for the electrons in MnO$_2$ and BaO to go to the CuO$_2$ layer.
The Cu-3d$_{z^2-r^2}$ + O-2p$_z$ charge difference regarding the interfacial charge transfer is illustrated in Appendix~\ref{int_chg_transf}. 

\subsection{\label{subsec:mag}Magnetism}

\begin{figure*}
\includegraphics[scale=0.55]{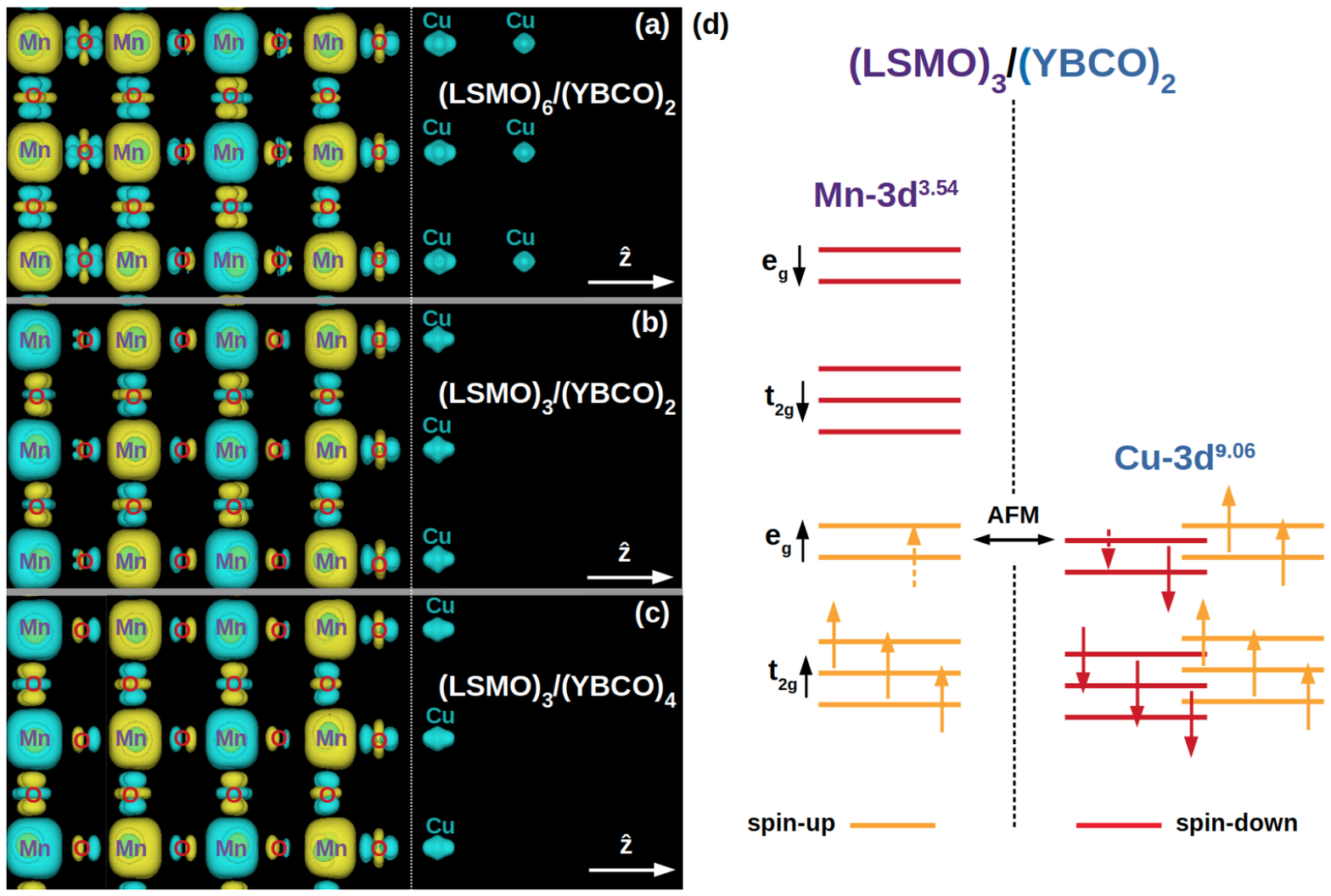}% Here is how to import EPS art
\caption{Spin densities $\Delta\rho=\rho_{\uparrow}-\rho_{\downarrow}$ isosurfaces for (a) (LSMO)$_6$/(YBCO)$_2$, (b) (LSMO)$_3$/(YBCO)$_2$ and (c) (LSMO)$_3$/(YBCO)$_4$ superlattices. The surfaces in yellow represent $\Delta\rho>0$, and the ones in blue represent $\Delta\rho<0$. Dotted lines indicate the interface position.  In (d) we depicted the Cu (Mn) 3$d$ occupancies and the hybridization responsible for the AFM coupling. Spin-up (spin-down) electrons are indicated by yellow (red) arrows. Dashed arrows indicate fractional orbital occupations.}
\label{fig:6}
\end{figure*}

Having established the structural distortions and charge transfer between YBCO and LSMO, we now explore the obtained DFT+U magnetic ground states for each superlattice.
In Fig.~\ref{fig:6} we display the spin density $\Delta \rho=\rho_{\uparrow}-\rho_{\downarrow}$ for (a) (LSMO)$_6$/(YBCO)$_2$, (b) (LSMO)$_3$/(YBCO)$_2$ and (c) (LSMO)$_3$/(YBCO)$_4$ superlattices, respectively. First, one can notice that the magnetization profile of LSMO in all these structures corresponds to in-plane FM coupled Mn-local moments with AFM ordering between planes, giving rise to the so-called magnetic ``dead layer". More importantly, we observe the formation of tiny Cu-centered local magnetic moments up to the second CuO$_2$ plane for (LSMO)$_6$/(YBCO)$_2$, whereas for (LSMO)$_3$/(YBCO)$_2$ and (LSMO)$_3$/(YBCO)$_4$ the induced local moments only appear in the interfacial CuO$_2$ plane. For the interfacial Cu atoms, we find tiny local moments of 0.04 $\mu_{B}$/Cu for (LSMO)$_6$/(YBCO)$_2$ and 0.03 $\mu_{B}$/Cu for (LSMO)$_3$/(YBCO)$_4$, as indicated in Fig.~\ref{fig:4}. For the (LSMO)$_3$/(YBCO)$_2$ superlattice the calculated local moments are the same as in (LSMO)$_6$/(YBCO)$_2$ (not shown). This is in reasonable agreement with the local moments reported experimentally~\cite{dead_layer}, where a value of 0.06 $\mu_{B}$/Cu was found. The AFM coupling between Cu and Mn atoms that we observe at the interfaces can be explained using the Goodenough-Kanamori-Anderson (GKA) rules for the superexchange interaction~\cite{gka,gka2}. As schematically depicted in Fig.~\ref{fig:6}(d), our calculations indicate that there is a net antiferromagnetic coupling due to the Cu$^{2+}$(3$d^9$)-Mn$^{3+}$(3$d^4$) hybridization, and a sizable orbital polarization concerning the 3$d_{x^2-y^2}$ orbital.

\begin{figure}
\includegraphics[scale=0.4]{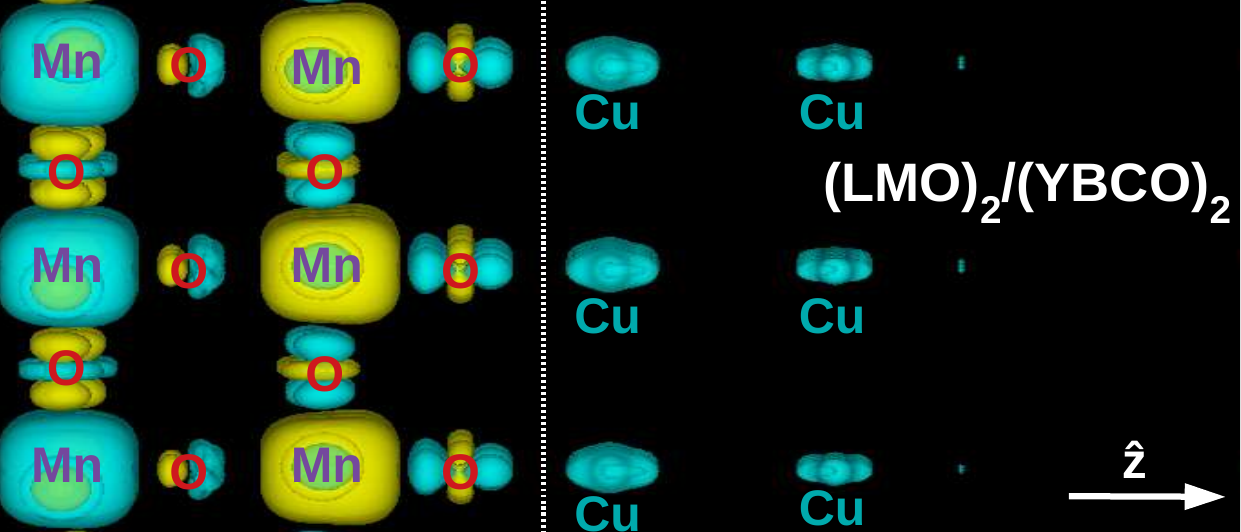}% Here is how to import EPS art
\caption{Spin densities $\Delta\rho=\rho_{\uparrow}-\rho_{\downarrow}$ isosurfaces for the (LMO)$_2$/(YBCO)$_2$ superlattice. The surfaces in yellow represent $\Delta\rho>0$, and the ones in blue represent $\Delta\rho<0$. Dotted lines indicate the interface position. }
\label{fig:7}
\end{figure}

For the undoped (LMO)$_2$/(YBCO)$_2$ superlattice (Fig.~\ref{fig:7}), the induced magnetic moments appear in the Cu-chain atoms near the interface, with values of 0.04 $\mu_{B}$/Cu. We see the formation of local magnetic moments up to the second Cu plane, indicating that without Sr-doping, the interfacial magnetic effect is stronger for smaller manganite thickness (not shown). In the case of the heterostructure (see Appendix~\ref{het}) the local moments are induced only in the first layer of Cu atoms, and we find tiny local moments of 0.04 $\mu_{B}$/Cu.

We also analyzed interfaces forced to be fully ferromagnetic, i.e. without the presence of the ``dead layer". These results are presented in Appendix~\ref{dl_mag}. We see that the charge transfer decreases, while the magnetic moments induced in the interfacial CuO$_2$ layers have a small increase of $0.02\mu_B$/Cu when LSMO is completely ferromagnetic.

On the right axes of Fig. \ref{fig:4} one can see the magnetization profile for the (LSMO)$_6$/(YBCO)$_2$ and (LSMO)$_3$/(YBCO)$_4$ superlattices. Due to the charge transfer, the Mn atoms close to the interface exhibit smaller values than most intern ones, reaching $3.30\mu_B$ in (LSMO)$_3$/(YBCO)$_2$ (not shown). By increasing the YBCO layer thickness, we observe a reduction of $0.11\mu_B$ in the magnitude of the first Mn local moment and of $0.04\mu_B$ in the second Mn atom for the (LSMO)$_3$/(YBCO)$_4$ superlattice. On the other hand, the increase in the number of LSMO unit cells leads to the opposite effect, where we find an increase of $0.04\mu_B$ in the first Mn magnetic moment and of $0.05\mu_B$ in the second Mn atom for the (LSMO)$_6$/(YBCO)$_2$ superlattice. The inner LSMO layers have a bulk-like magnetism. It is noteworthy to emphasize that our calculations indicate a decoupling between the induced charge transfers and the local moments in the cuprate, as we explain in the following. For instance, one can notice that in the case of (LSMO)$_3$/(YBCO)$_4$, the induced local moments appear only in the interfacial CuO$_2$ planes, whereas in (LSMO)$_6$/(YBCO)$_2$ it extends to the second CuO$_2$ layer. For the former, the charge transfer occurs up to the Cu5 plane near the bulk-like Cu-chain layer. 
Another interesting fact is that the interfacial MnO$_2$ plane, even with a larger occupation than in bulk, has a smaller magnetic moment, which can indicate a change in the in-plane magnetic coupling that leads the system closer to an AFM configuration, reducing the strength of the Hund's coupling. 

To summarize, our results indicate that the penetration depth of interfacial local moments is controlled by the manganite thickness and Sr concentration. 
More importantly, these local moments give rise to an in-plane ferromagnetic ordering within the interfacial CuO$_2$ planes, which competes against the formation of singlet Cooper pairs and may explain the attenuation of $T_c$ in heterostructures of manganites/cuprates. It is also important to mention that strain-induced deoxygenation plays an important role in the suppression of the superconducting phase of YBCO in other heterostructures~\cite{zhang2023}. 

\subsection{\label{int_hyb} Mn-O-Cu Hybridization}

To study the hybridization between the interfacial Mn-O-Cu states, we calculated the projected density of states (PDOS) for the (LSMO)$_3$/(YBCO)$_2$ superlattice. In particular, we focused on the Cu-3$d_{z^2-r^2}$, Mn-3$d_{z^2-r^2}$ and O-2$p_z$ contributions for the electronic states around the Fermi energy. Due to the charge transfer, we observe that the Cu-3$d_{z^2-r^2}$ states in the (LSMO)$_3$/(YBCO)$_2$ superlattice downshift (not shown) by around 0.5 eV in comparison to the same states in the isolated bulk YBCO.

\begin{figure}
\includegraphics[scale=0.5]{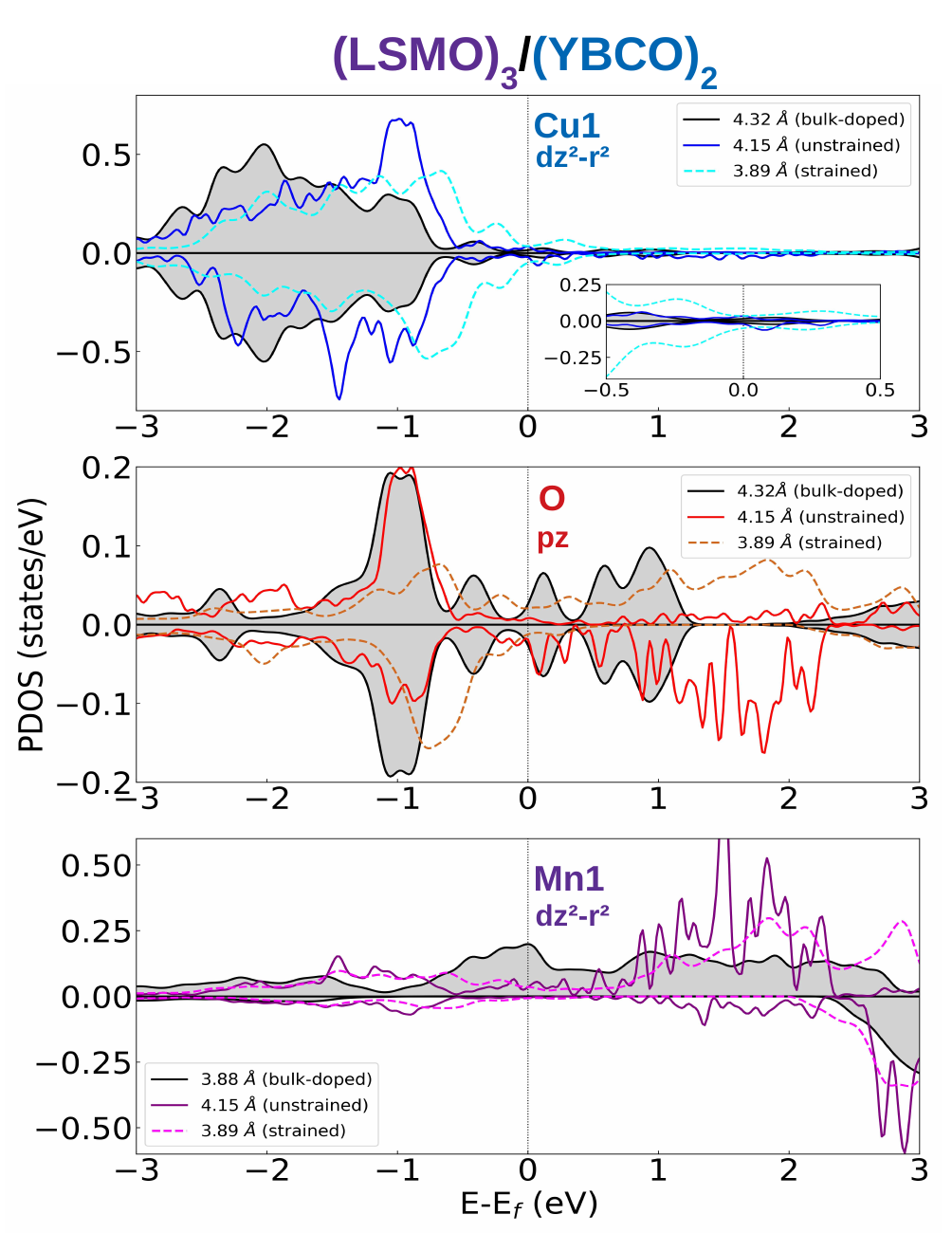}% Here is how to import EPS art
\caption{PDOS in the orbitals associated to the Mn-O-Cu bond for the (LSMO)$_3$/(YBCO)$_2$ superlattice with (dashed lines) and without (continuous lines) $z$-axis strain, compared to the respective bulk-doped values (in black). Positive (negative) values signify spin-up (spin-down) contributions. The orbitals chosen were 3d$_{z^2-r^2}$ for Cu (in blue) and Mn (in purple) and 2p$_z$ for O (in red). The inset in the first panel is a zoom in the -0.5 eV to 0.5 eV energy window. The values in \AA{} in the legends indicate the Mn-O-Cu distances for the two superstructures considered by us, as well as the bulk-doped YBCO (LSMO) Cu-Cu (Mn-Mn) distance in $z$-axis direction.}
\label{est_ele_int}
\end{figure}

To disentangle the effects of the hybridization from the induced band shift due to charge transfer, we show in Fig.~\ref{est_ele_int} the calculated PDOS for the (LSMO)$_3$/(YBCO)$_2$ superlattice and the ones calculated for electron-doped (hole-doped) bulk YBCO (LSMO). The doping values were chosen in order to simulate the same amount of charge transfer observed in the (LSMO)$_3$/(YBCO)$_2$ superlattice. In addition, we calculate the electronic structure of (LSMO)$_3$/(YBCO)$_2$ under a compressive strain of 5\% in the $z$-axis parameter to reduce the distance between the interfacial Mn-Cu atoms.
In this case, our unstrained structure indicates a distance of 4.15 \AA{} between the interfacial Mn and Cu atoms, while the strained structure indicates a distance of 3.89 \AA{}. For comparison, in the doped structures of bulk YBCO (LSMO), the Cu-Cu (Mn-Mn) distance in $z$-axis direction is 4.32 \AA{} (3.88 \AA{}).

As can be noticed in Fig.~\ref{est_ele_int}, the electronic states of interfacial atoms in the unstrained (LSMO)$_3$/(YBCO)$_2$ superlattice (Mn-O-Cu distance of 4.15 \AA{}) exhibit an energy alignment around -1 eV, indicating an out-of-plane orbital hybridization. The same feature is observed between 1 and 2 eV due to the hybridization of Mn-3d$_{z^2}$ and O-2p$_z$. Despite this sizable hybridization, we do not observe a substantial orbital reconstruction concerning the interfacial Cu-3$d_{z^2-r^2}$ states in this case. 
On the other hand, under compressive strain in the $z$ direction, we induce an increase of the Mn-O-Cu hybridization and an upshift of the Cu-3$d_{z^2-r^2}$ states towards the Fermi energy. As a result, the number of holes in the Cu-3$d$ states would be redistributed between the Cu-3$d_{x^2-y^2}$ and 3$d_{z^2-r^2}$, leading to a sizable orbital reconstruction.

\section{\label{sec:conc}Conclusions}

In summary, we performed DFT+U calculations to investigate the interfacial effects in LSMO/YBCO superlattices. 
We find that the presence of epitaxial strain and the interface leads to the reduction of buckling parameters ($b$ and $\phi$) in the YBCO layer, compared with bulk values, which is beneficial to stabilizing the superconducting phase. We also observe that the interfacial MnO$_2$ layers in LSMO transfer electrons to the first CuO$_2$ planes in YBCO, which increases the Cu-3$d$ occupancy. However, the change in Cu-3$d$ valence only slightly depends on the LSMO thickness.
Additionally, we find induced local moments in the interfacial CuO$_2$ planes, which order in an in-plane ferromagnetic state that competes against superconductivity. These local moments are decoupled from the charge transfer and, according to our calculations, appear mainly due to the Mn 3d-O 2p-Cu 3d hybridization, which can give rise to an orbital reconstruction for the interfacial Cu-3$d$ states. 
From the manganite side, we observe a reduction in the magnetic moment of the interfacial Mn atoms, which can be connected to a decrease in the FM coupling strength caused by the Mn electron doping, driving the system in the direction of an AFM phase. The formation of a magnetic "dead layer" can be seen as an AFM out-of-plane coupling in the first LSMO layers caused by the structural distortions near the interface, which gives rise to a vanishing magnetization around 3\AA{} from the interface. 
Our findings provide the essential ingredients to understand complex phenomena, including suppression of the superconductivity, in YBCO/LSMO superlattices.

\begin{acknowledgments}

The authors acknowledge the financial support from the Brazilian agencies CNPq (in particular Grants 402919/2021-1 and INCT-IQ 465469/2014-0), CAPES, FAPEMIG, and the computational centers: National Laboratory for Scientific Computing (LNCC/MCTI, Brazil), for providing HPC resources of the SDumont supercomputer (URL: http://sdumont.lncc.br),  CENAPAD-SP, and CESUP-UFRGS.

\end{acknowledgments}

% ---
% Inicia os apêndices
% ---
\appendix

\section{LMO/YBCO heterostructure}\label{het}

\begin{figure}
\includegraphics[scale=0.26]{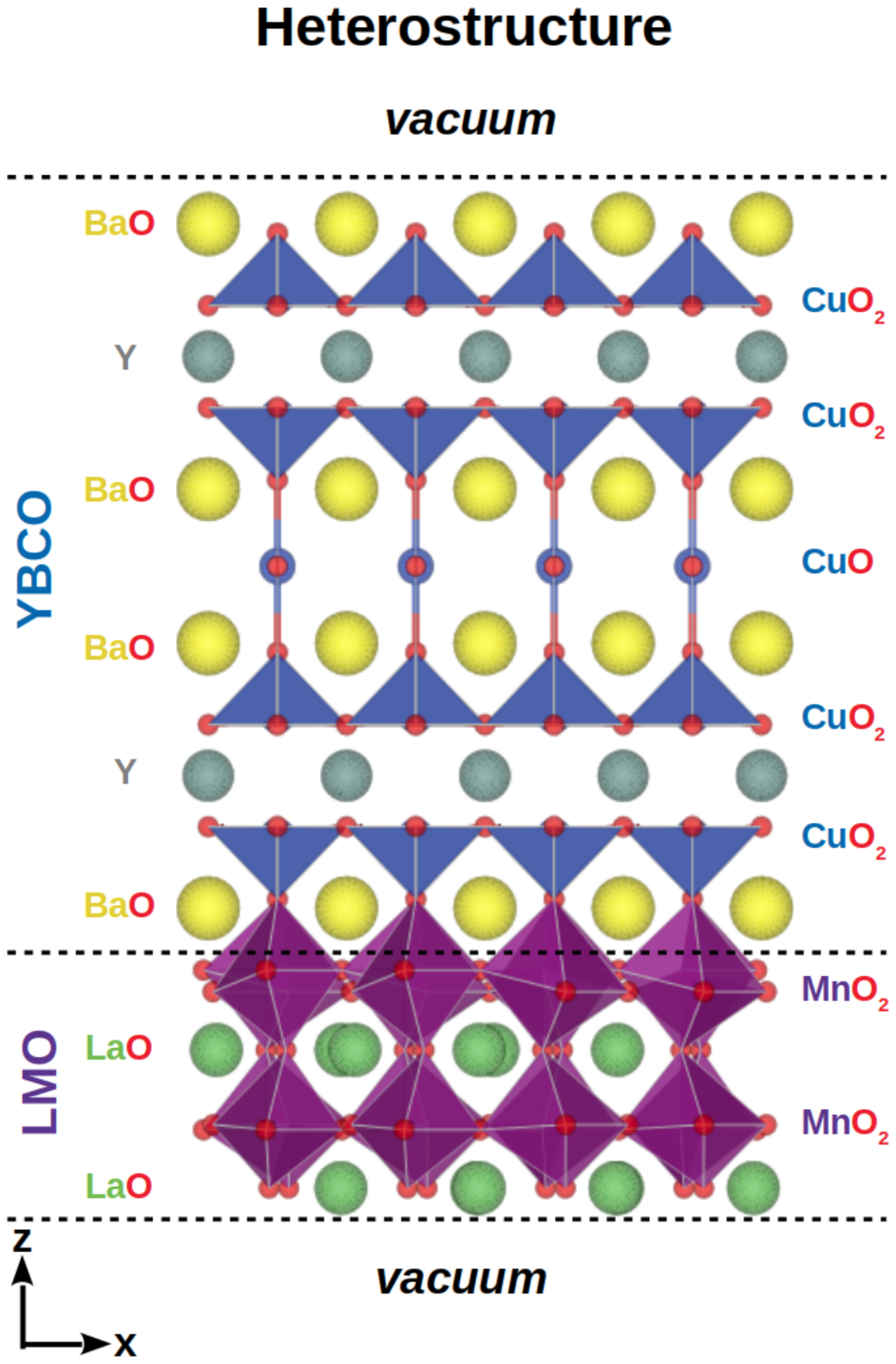}% Here is how to import EPS art
\caption{Structural model of the LaMnO$_3$ (2 u.c.)/YBa$_2$Cu$_3$O$_7$ (2 u.c.) heterostructure. The dashed lines indicate the interface position.}
\label{fig:s1}
\end{figure}

\begin{figure}
\includegraphics[scale=0.33]{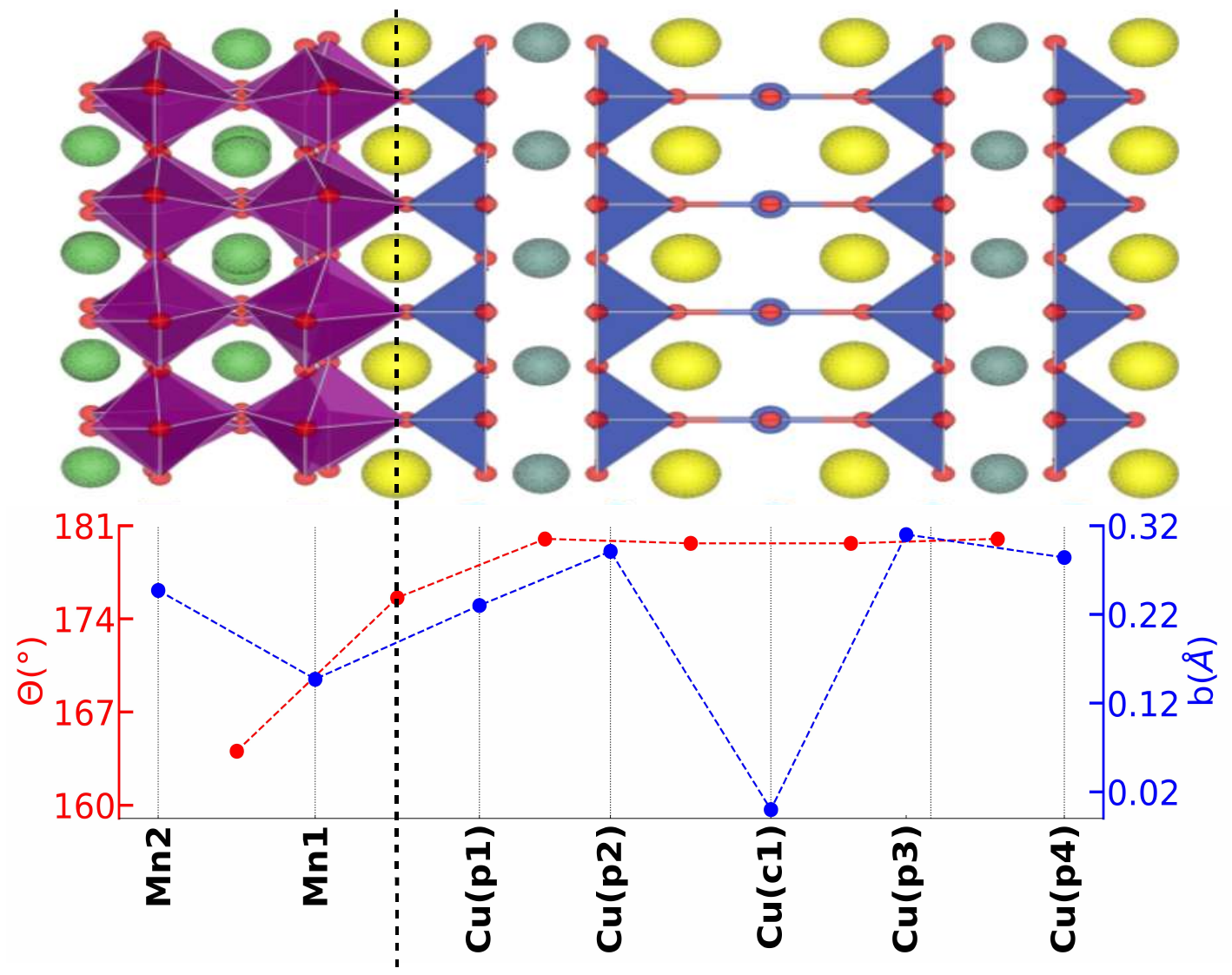}% Here is how to import EPS art
\caption{Out-of-plane transition metal-oxygen-transition metal bond angles $\theta$ and in-plane buckling $b$ for the heterostructure. The dashed lines indicate the interface's location.}
\label{fig:s2}
\end{figure}

For the LaMnO$_3$ (2 u.c.)/YBa$_2$Cu$_3$O$_7$ (2 u.c.) heterostructure, we considered realistic thin-film geometries using similar supercells with BaO termination and vacuum separation of 20 \AA{} (see Fig.~\ref{fig:s1}). In this case, we only used LMO. The relaxed heterostructure is shown in Fig.~\ref{fig:s2}. We find that the buckling distance $b$ is slightly reduced for the CuO$_2$ planes at the interface (0.061 \AA{}). According to our results, the reduction of $b$ is accompanied by an average reduction of the buckling angle $\phi$ of around 2.88$^\circ$ in the heterostructure. As can be seen in Fig.~\ref{fig:s2} (red lines), $\theta$ near the interface approaches values corresponding to large doping, suggesting an AFM coupling of MnO$_2$ layers (magnetic ``dead layer") close to the interface.

\begin{figure}
\includegraphics[scale=0.27]{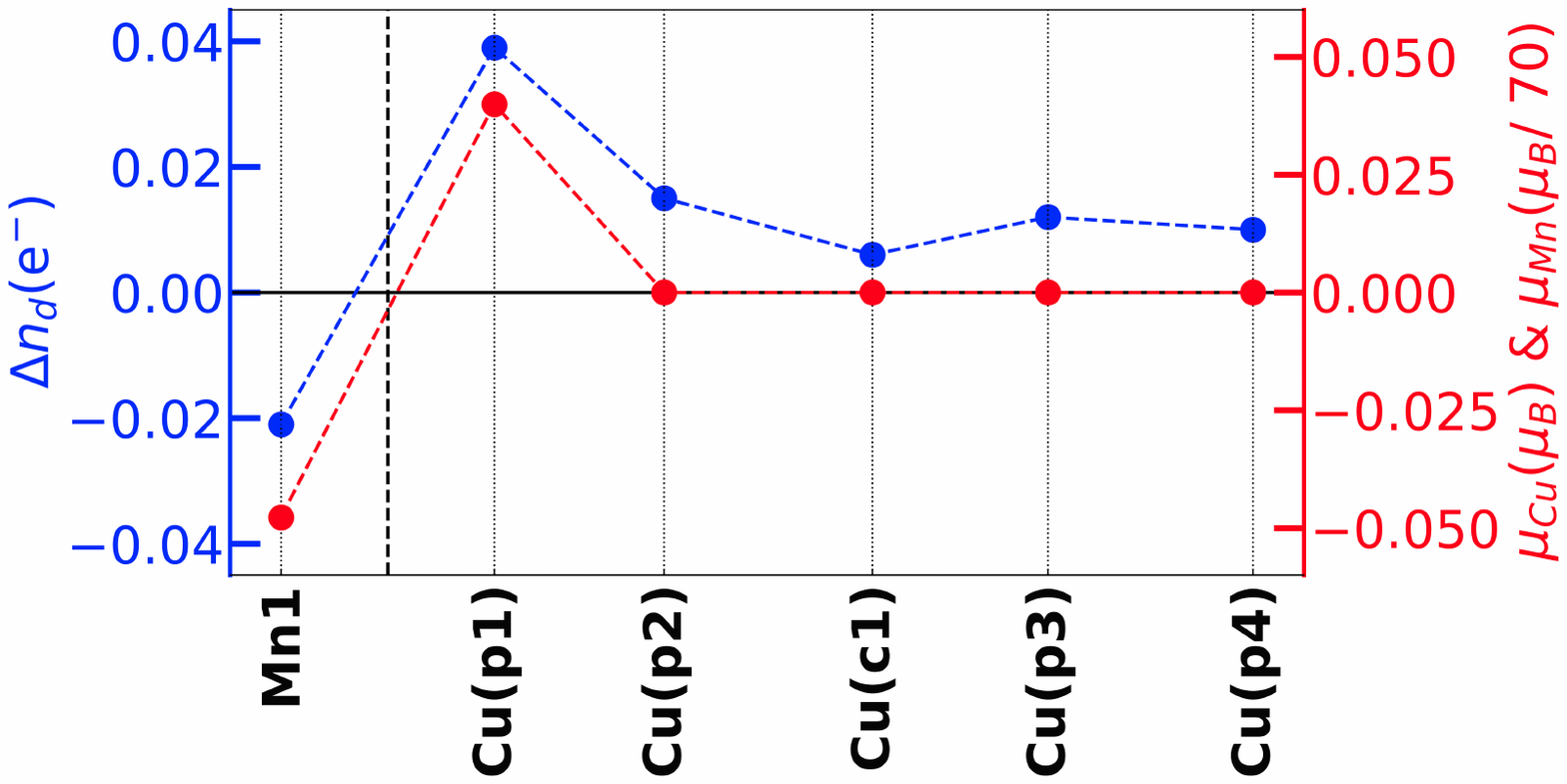}% Here is how to import EPS art
\caption{3d-orbital charge difference $\Delta_n$ (blue) and magnetic moments $\mu$ (red) for the TM along the heterostructure. The dashed black lines indicates the interfaces. For better visualization, the Mn moments were reduced by a factor of 70. $\Delta_n$ was calculated using the difference between the interface and bulk TM occupations.}
\label{fig:s3}
\end{figure}

In Fig.~\ref{fig:s3} we show the variation of 3$d$ occupancies (in blue), $\Delta n_d = n_{Hetero} - n_{bulk}$, for Mn and Cu atoms for the heterostructure. In this structure, the Cu-3$d$ interfacial occupancies increase by 0.014, which should drive the CuO$_2$ planes away from the optimal regime (towards the antiferromagnetic phase). On the right panels of Fig.~\ref{fig:s3} one can see the magnetization profile for the heterostructure, with a decay of the magnetic moment of the Mn atoms close to the interface and a small magnetic moment induced in the first layer Cu atoms, with the same order as reported in XAS experiments.

\begin{figure}
\includegraphics[scale=0.4]{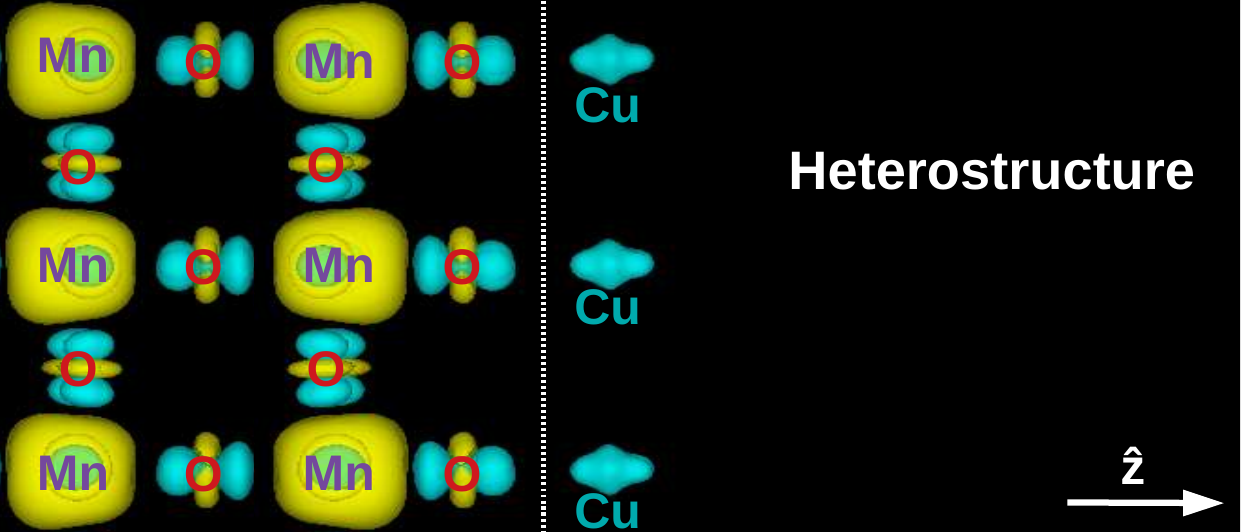}% Here is how to import EPS art
\caption{Spin densities $\Delta\rho=\rho_{\uparrow}-\rho_{\downarrow}$ isosurfaces for the heterostructure. Dotted lines indicate the interface position, the surfaces in yellow represents $\Delta\rho>0$ and the ones in blue represents $\Delta\rho<0$.}
\label{fig:s4}
\end{figure}

In Fig.~\ref{fig:s4} we display the spin density $\Delta \rho=\rho_{\uparrow}-\rho_{\downarrow}$ for the heterostructure, which exhibits local moments only in the interfacial CuO$_2$ plane. For the interfacial Cu atoms, we find tiny local moments of 0.04 $\mu_{B}$/Cu, in reasonable agreement with the local moments reported experimentally.

\section{Strain effects in T$_c$}\label{st_tc}

\begin{table}[hbt!]
\centering
\begin{tabular}{P{2.7cm} P{1.2cm} P{1.4cm} P{1.2cm} P{1.2cm}}
\hline
\hline
 Structure & $\epsilon_z$ (\%) & Cu$^1$-Cu$^2$ (\AA) & Cu-AO (\AA) & Ba-AO (\AA) \\
 \hline
 & & & & \\
 YBCO & +0.0 & 4.14 & 2.25 & 0.24 \\
 & & & & \\
 (LMO)$_2$/(YBCO)$_2$ & +1.9 & 3.31 & 2.25 & 0.22 \\
 & & & & \\
 (LSMO)$_3$/(YBCO)$_2$ & +0.9 & 3.30 & 2.30 & 0.23 \\
 & & & & \\
 Heterostructure & +0.7 & 3.34 & 2.35 & 0.33 \\
 & & & & \\
 (LSMO)$_6$/(YBCO)$_2$ & -0.2 & 3.34 & 2.36 & 0.26 \\
 & & & & \\
 (LSMO)$_3$/(YBCO)$_4$ & -0.6 & 3.34 & 2.38 & 0.27 \\
 & & & & \\
 (LMO)$_3$/(YBCO)$_2$ & -0.8 & 3.34 & 2.45 & 0.38 \\
\hline
\hline
\end{tabular}
\caption{$z$-axis strain $\epsilon_z=-\epsilon_{33}$ (first column), interlayer Cu-Cu spacing (second column), copper to apical oxygen (AO) distance (third column), and vertical component of Ba to AO distance (fourth column) for the interfaces. We also report distances for the pristine material (first row) confined in the STO parameters.}
\label{tab:s1}
\end{table}

In Table~\ref{tab:s1} we show the $z$-axis strain $\epsilon_z$ values and some interatomic distances across the interfaces. A recent theoretical study~\cite{strain_tc} reported a connection between $\epsilon_z$ and these distances with the superconducting critical temperature $T_c$ for bulk YBCO. The authors find a reduction of $T_c$ up to 50$\%$ when strain goes from 0.0$\%$ to 4.0$\%$, with a reduction in Cu(p1)-Cu(p2), Cu-apical oxygen (AO) and Ba-AO distances. Our results for positive strain (compressive) agree qualitatively with the tendencies for YBCO, with the (LSMO)$_3$/(YBCO)$_2$ superstructure having distance values close to the 4.0$\%$ strained YBCO, indicating a possible reduction in $T_c$. It is worth mentioning this is the same $m/n$ ratio in which the superconducting-to-insulator transition occurs for LCMO/YBCO interfaces. The heterostructure distances are somewhere between 0.0$\%$ and 4.0$\%$ YBCO bulk values, indicating a possible reduction in $T_c$. The interfaces with negative strain cannot be compared because of a lack of comparative data.

\section{Interfacial charge transfer}\label{int_chg_transf}

\begin{figure}[hbt!]
\includegraphics[scale=0.27]{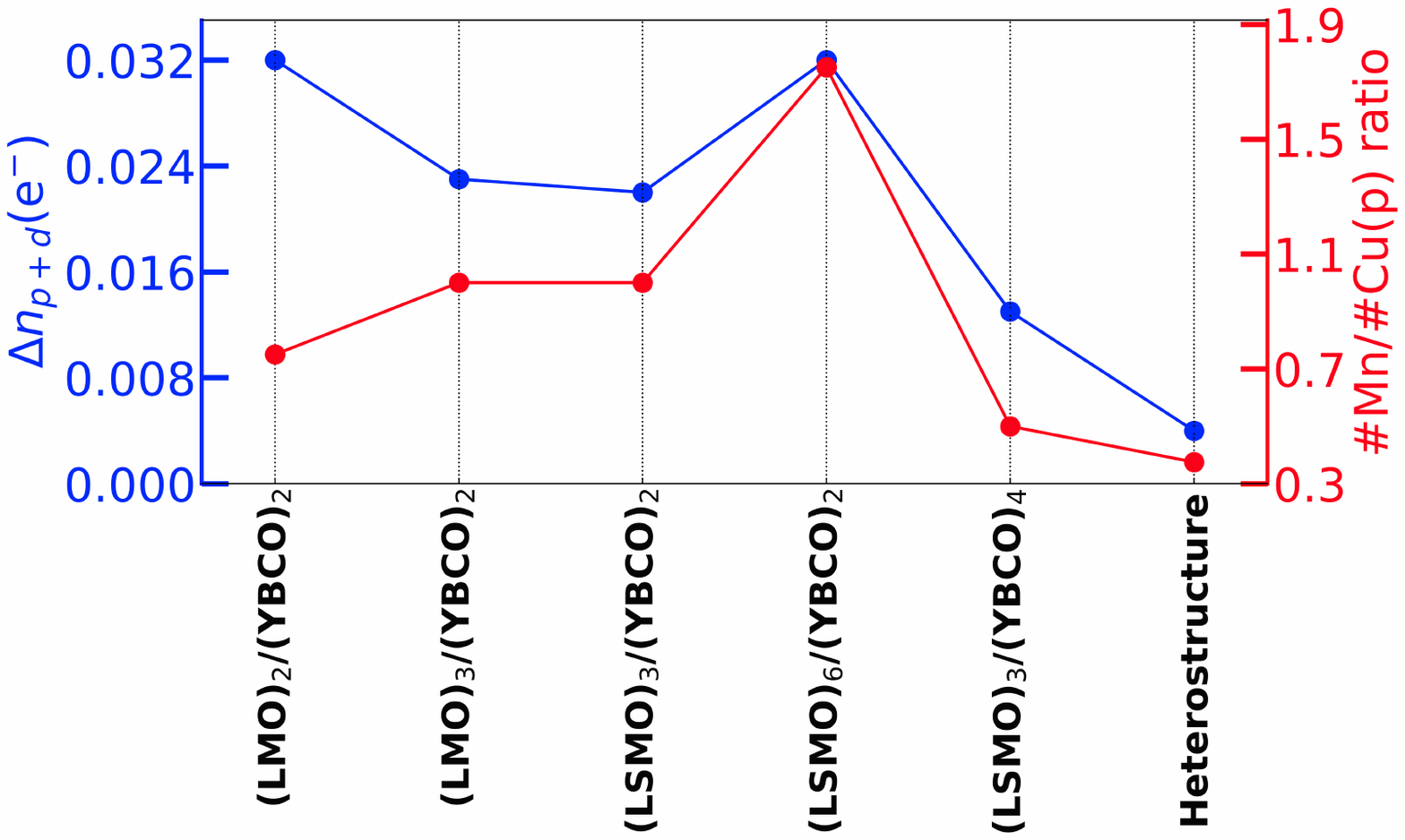}% Here is how to import EPS art
\caption{Cu-3d$_{z^2-r^2}$ + O-2p$_z$ charge difference $\Delta_n$ (in blue) and (\# MnO$_2$)/(\# CuO$_2$) ratio (in red) for the interfaces. $\Delta_n$ was calculated using the difference between the interface and bulk occupation values. ``Hetero'' is associated to the LMO (2 u.c)/YBCO (2 u.c.) heterostructure.}
\label{fig:s5}
\end{figure}

Fig.~\ref{fig:s5} indicates the Cu-3d$_{z^2-r^2}$ + O-2p$_z$ charge difference $\Delta_n$ (in blue), together with the (\# MnO$_2$)/(\# CuO$_2$) ratio (in red), for all the superstructures. We observe that (LSMO)$_6$/(YBCO)$_2$ and (LMO)$_3$/(YBCO)$_2$ have the largest occupation values and that increasing the YBCO thickness ((LSMO)$_3$/(YBCO)$_2$$\rightarrow$(LSMO)$_3$/(YBCO)$_4$) decreases the total charge transferred. Increasing the LSMO thickness ((LSMO)$_3$/(YBCO)$_2$$\rightarrow$(LSMO)$_6$/(YBCO)$_2$) has the opposite effect, increasing the charge in Cu-3d$_{z^2-r^2}$ and O-2p$_z$ orbitals. This behavior is the same for the (\# MnO$_2$)/(\# CuO$_2$) ratio, except for (LMO)$_2$/(YBCO)$_2$, indicating that the proportion between Mn and Cu atoms can be used to control the doping level of the CuO$_2$ planes close to the interface and affect their electronic structure.

\section{Magnetic ``Dead layer" versus Ferromagnetism}\label{dl_mag}

\begin{figure*}[hbt!]
\includegraphics[scale=0.4]{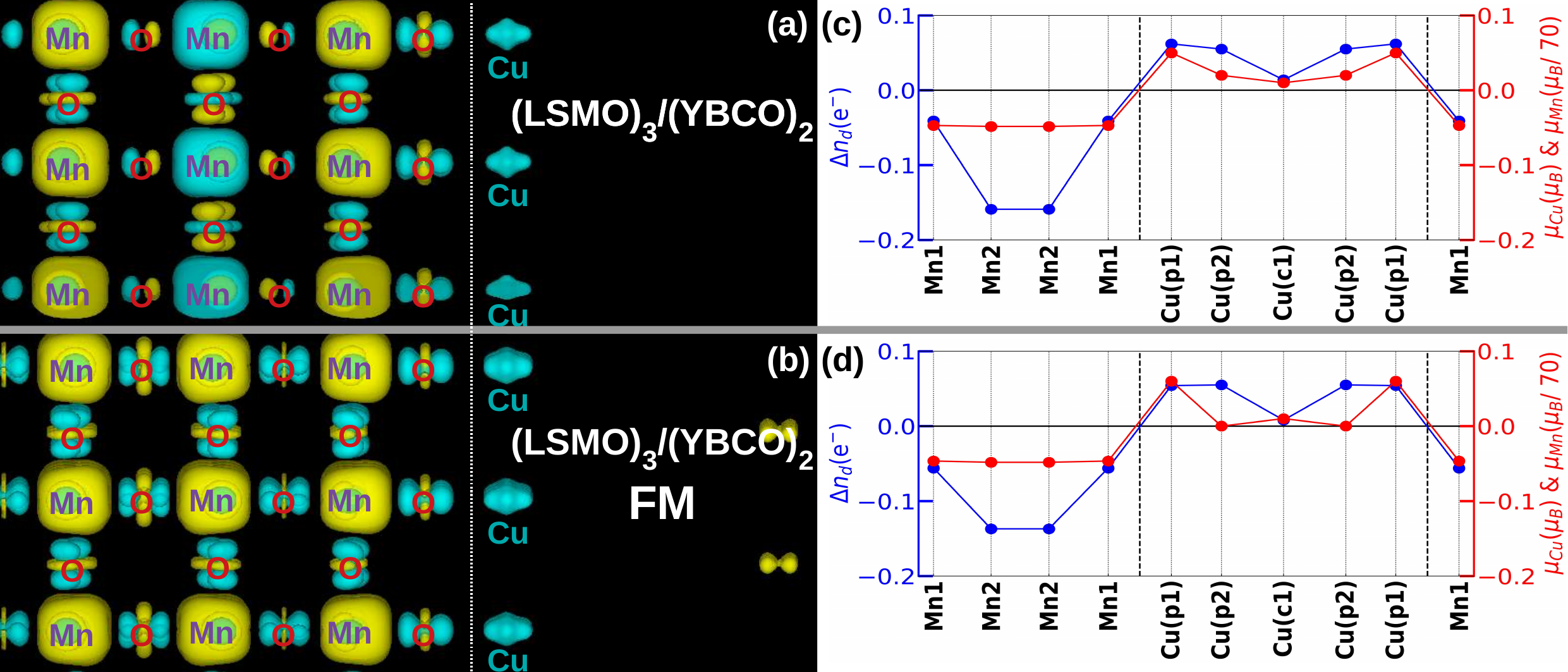}% Here is how to import EPS art
\caption{Spin densities $\Delta\rho=\rho_{\uparrow}-\rho_{\downarrow}$ isosurfaces for (LSMO)$_3$/(YBCO)$_2$ (a) with ``dead layer" and (b) with LSMO full FM. (c) and (d) are the respective charge difference $\Delta_n$ (in blue) and magnetic moment $\mu$ (in red). Dotted lines indicate the interface position, the surfaces in yellow represents $\Delta\rho>0$ and the ones in blue represents $\Delta\rho<0$. For better visualization, the Mn moments were reduced by a factor of 70.}
\label{fig:s6}
\end{figure*}

In Fig.~\ref{fig:s6} we show a comparison between the charge and magnetic profiles of (LSMO)$_3$/(YBCO)$_2$ with ``dead layer" and with full ferromagnetic (FM) LSMO. We observe that the charge transfer across the interface is reduced when LSMO is FM, and the local magnetic moment is finite only for the first plane, but with a larger value compared to the ``dead layer".

\nocite{*}

\bibliography{main}% Produces the bibliography via BibTeX.

\end{document}